\documentclass{myarticle} 
\usepackage{nips11submit_e,times}
\usepackage{amsmath,amssymb}
\usepackage{dsfont}
\usepackage{subfigure}
\usepackage{graphics}
\usepackage{epsfig}
\usepackage{multirow}
\usepackage{color}
\usepackage{times}

\usepackage{amsthm}

\newtheorem{defn}{Definition}
\newtheorem{cor}{Corollary}
\newtheorem{prop}{Proposition}

\title{Generalized Beta Mixtures of Gaussians}

\author{
Artin Armagan  \\
Dept. of Statistical Science\\
Duke University\\
Durham, NC 27708 \\
\texttt{artin@stat.duke.edu} \\
\And
David B. Dunson  \\
Dept. of Statistical Science\\
Duke University\\
Durham, NC 27708 \\
\texttt{dunson@stat.duke.edu} \\
\And
Merlise Clyde  \\
Dept. of Statistical Science\\
Duke University\\
Durham, NC 27708 \\
\texttt{clyde@stat.duke.edu} \\
}

\nipsfinalcopy 

\begin{document}

\maketitle

\begin{abstract}
In recent years, a rich variety of shrinkage priors have been proposed that have great promise in addressing massive regression problems.  In general, these new priors can be expressed as scale mixtures of normals, but have more complex forms and better properties than traditional Cauchy and double exponential priors. We first propose a new class of normal scale mixtures through a novel generalized beta distribution that encompasses many interesting priors as special cases.  This encompassing framework should prove useful in comparing competing priors, considering properties and revealing close connections. We then develop a class of variational Bayes approximations through the new hierarchy presented that will scale more efficiently to the types of truly massive data sets that are now encountered routinely.
\end{abstract}

\section{Introduction} \label{Introduction}
Penalized likelihood estimation has evolved into a major area of research, with $\ell_1$\cite{tibshirani1996} and other regularization penalties now used routinely in a rich variety of domains. Often minimizing a loss function subject to a regularization penalty leads to an estimator that has a Bayesian interpretation as the mode of a posterior distribution \cite{figo03,griffin2007,armagan09a,armagan11barxiv}, with different prior distributions inducing different penalties. For example, it is well known that Gaussian priors induce $\ell_2$ penalties, while double exponential priors induce $\ell_1$ penalties \cite{figo03,park2008,hans2009,armagan09a}. Viewing massive-dimensional parameter learning and prediction problems from a Bayesian perspective naturally leads one to design new priors that have substantial advantages over the simple normal or double exponential choices and that induce rich new families of penalties.  For example, in high-dimensional settings it is often appealing to have a prior that is concentrated at zero, favoring strong shrinkage of small signals and potentially a sparse estimator, while having heavy tails to avoid over-shrinkage of the larger signals.  The Gaussian and double exponential priors are insufficiently flexible in having a single scale parameter and relatively light tails; in order to shrink many small signals strongly towards zero, the double exponential must be concentrated near zero and hence will over-shrink signals not close to zero.  This phenomenon has motivated a rich variety of new priors such as the \emph{normal-exponential-gamma}, the \emph{horseshoe} and the \emph{generalized double Pareto} \cite{griffin2007,hoggart08,armagan09a,Carvalho2009,polson2009,carvalho2010,griffin2010,armagan11barxiv}.

An alternative and widely applied Bayesian framework relies on variable selection priors and Bayesian model selection/averaging \cite{mitchell1988,george93,johnstone2004,ishwaran2005}. Under such approaches the prior is a mixture of a mass at zero, corresponding to the coefficients to be set equal to zero and hence excluded from the model, and a continuous distribution, providing a prior for the size of the non-zero signals. This paradigm is very appealing in fully accounting for uncertainty in parameter learning and the unknown sparsity structure through a probabilistic framework.  One obtains a posterior distribution over the model space corresponding to all possible subsets of predictors, and one can use this posterior for model-averaged predictions that take into account uncertainty in subset selection and to obtain marginal inclusion probabilities for each predictor providing a weight of evidence that a specific signal is non-zero allowing for uncertainty in the other signals to be included.  Unfortunately, the computational complexity is exponential in the number of candidate predictors ($2^p$ with $p$ the number of predictors). 

Some recently proposed continuous shrinkage priors may be considered competitors to the conventional mixture priors \cite{ishwaran2005,Carvalho2009,carvalho2010,griffin2010} yielding computationally attractive alternatives to Bayesian model averaging. Continuous shrinkage priors lead to several advantages. The ones represented as scale mixtures of Gaussians allow conjugate block updating of the regression coefficients in linear models and hence lead to substantial improvements in Markov chain Monte Carlo (MCMC) efficiency through more rapid mixing and convergence rates. Under certain conditions these will also yield sparse estimates, if desired, via maximum a posteriori ($\mbox{MAP}$) estimation and approximate inferences via variational approaches \cite{jordan99,tipping01,bishop01,figo03,griffin2007,armagan09a,armagan11barxiv}.  

The class of priors that we consider in this paper encompasses many interesting priors as special cases and reveals interesting connections among different hierarchical formulations. Exploiting an equivalent conjugate hierarchy of this class of priors, we develop a class of variational Bayes approximations that can scale up to truly massive data sets. This conjugate hierarchy also allows for conjugate modeling of some previously proposed priors which have some rather complex yet advantageous forms and facilitates straightforward computation via Gibbs sampling. We also argue intuitively that by adjusting a global shrinkage parameter that controls the overall sparsity level, we may control the number of non-zero parameters to be estimated, enhancing results, if there is an underlying sparse structure. This global shrinkage parameter is inherent to the structure of the priors we discuss as in  \cite{Carvalho2009,carvalho2010} with close connections to the conventional variable selection priors. 

\section{Background}

We provide a brief background on shrinkage priors focusing primarily on the priors studied by \cite{Carvalho2009,carvalho2010} and \cite{griffin2007,griffin2010} as well as the Strawderman-Berger (SB) prior \cite{carvalho2010}. These priors possess some very appealing properties in contrast to the double exponential prior which leads to the Bayesian lasso \cite{park2008,hans2009}. They may be much heavier-tailed, biasing large signals less drastically while shrinking noise-like signals heavily towards zero. In particular, the priors by \cite{Carvalho2009,carvalho2010}, along with the Strawderman-Berger prior \cite{carvalho2010}, have a very interesting and intuitive representation later given in (\ref{horseshoe2}), yet, are not formed in a conjugate manner potentially leading to analytical and computational complexity.   

\cite{Carvalho2009,carvalho2010} propose a useful class of priors for the estimation of multiple means. Suppose a $p$-dimensional vector $\mathbf{y}|\boldsymbol\theta\sim \mathcal{N}(\boldsymbol\theta,\mathbf{I})$ is observed. The independent hierarchical prior for $\theta_{j}$ is given by
\begin{equation}
\theta_{j}|\tau_{j}\sim \mathcal{N}(0,\tau_{j}), \ \ \tau_{j}^{1/2}\sim \mathcal{C}^{+}(0,\phi^{1/2}),
\label{horseshoe1}
\end{equation}
for $j=1,\dots,p$, where $\mathcal{N}(\mu,\nu)$ denotes a normal distribution with mean $\mu$ and variance $\nu$ and $\mathcal{C}^{+}(0,s)$ denotes a half-Cauchy distribution on $\Re^{+}$ with scale parameter $s$. With an appropriate transformation $\rho_{j}=1/(1+\tau_{j})$, this hierarchy also can be represented as
\begin{equation}
\theta_{j}|\rho_{j}\sim \mathcal{N}(0,1/\rho_{j}-1), \ \ \pi(\rho_{j}|\phi)\propto\rho_{j}^{-1/2}(1-\rho_{j})^{-1/2}\frac{1}{1+(\phi-1)\rho_{j}}.
\label{horseshoe2}
\end{equation}
A special case where $\phi=1$ leads to $\rho_{j}\sim \mathcal{B}(1/2,1/2)$ (beta distribution) where the name of the prior arises, \emph{horseshoe} (HS) \cite{Carvalho2009,carvalho2010}. Here $\rho_{j}$s are referred to as the \emph{shrinkage coefficients} as they determine the magnitude with which $\theta_{j}$s are pulled toward zero. A prior of the form $\rho_{j} \sim \mathcal{B}(1/2,1/2)$ is natural to consider in the estimation of a signal $\theta_{j}$ as this yields a very desirable behavior both at the tails and in the neighborhood of zero. That is, the resulting prior has heavy-tails as well as being unbounded at zero which creates a strong pull towards zero for those values close to zero.  \cite{carvalho2010} further discuss priors of the form $\rho_{j}\sim \mathcal{B}(a,b)$ for $a>0$, $b>0$ to elaborate more on their focus on the choice $a=b=1/2$. A similar formulation dates back to \cite{strawderman1971}. \cite{carvalho2010} refer to the prior of the form $\rho_{j} \sim \mathcal{B}(1,1/2)$ as the Strawderman-Berger prior due to \cite{strawderman1971} and \cite{berger1980}. The same hierarchical prior is also referred to as the quasi-Cauchy prior in \cite{johnstone2004}. Hence, the tail behavior of the Strawderman-Berger prior remains similar to the horseshoe (when $\phi=1$), while the behavior around the origin changes. The hierarchy in (\ref{horseshoe2}) is much more intuitive than the one in (\ref{horseshoe1}) as it explicitly reveals the behavior of the resulting marginal prior on $\theta_{j}$. This intuitive representation makes these hierarchical priors interesting despite their relatively complex forms. On the other hand, what the prior in (\ref{horseshoe1}) or (\ref{horseshoe2}) lacks is a more trivial hierarchy that yields recognizable conditional posteriors in linear models.

\cite{griffin2007,griffin2010} consider the normal-exponential-gamma (NEG) and normal-gamma (NG) priors respectively which are formed in a conjugate manner yet lack the intuition the Strawderman-Berger and horseshoe priors provide in terms of the behavior of the density around the origin and at the tails. Hence the implementation of these priors may be more user-friendly but they are very implicit in how they behave. In what follows we will see that these two forms are not far from one another. In fact, we may unite these two distinct hierarchical formulations under the same class of priors through a generalized beta distribution and the proposed equivalence of hierarchies in the following section. This is rather important to be able to compare the behavior of priors emerging from different hierarchical formulations. Furthermore, this equivalence in the hierarchies will allow for a straightforward Gibbs sampling update in posterior inference, as well as making variational approximations possible in linear models.

\section{Equivalence of Hierarchies via a Generalized Beta Distribution} \label{prior}
In this section we propose a generalization of the beta distribution to form a flexible class of scale mixtures of normals with very appealing behavior. We then formulate our hierarchical prior in a conjugate manner and reveal similarities and connections to the priors given in  \cite{johnstone2004,griffin2007,griffin2010,Carvalho2009,carvalho2010}. As the name \emph{generalized beta} has previously been used, we refer to our generalization as the \emph{three-parameter beta} (TPB) distribution. 

In the forthcoming text $\Gamma(.)$ denotes the gamma function, $\mathcal{G}(\mu,\nu)$ denotes a gamma distribution with shape and rate parameters $\mu$ and $\nu$,  $\mathcal{W}(\nu,S)$ denotes a Wishart distribution with $\nu$ degrees of freedom and scale matrix $S$, $\mathcal{U}(\alpha_1,\alpha_2)$ denotes a uniform distribution over $(\alpha_1,\alpha_2)$, $\mathcal{GIG}(\mu,\nu,\xi)$ denotes a generalized inverse Gaussian distribution with density function $(\nu/\xi)^{\mu/2}\{2\mbox{K}_{\mu}(\sqrt{\nu\xi})\}^{-1}x^{\mu-1}\exp\{(\nu x+\xi/x)/2\}$, and $\mbox{K}_{\mu}(.)$ is a modified Bessel function of the second kind.

\begin{defn}
The three-parameter beta (TPB) distribution for a random variable $X$ is defined by the density function
\begin{equation}
f(x;a,b,\phi)=\frac{\Gamma(a+b)}{\Gamma(a)\Gamma(b)}\phi^{b}x^{b-1}(1-x)^{a-1}\left\{1+(\phi-1)x\right\}^{-(a+b)},
\label{betalike}
\end{equation}
for $0<x<1$, $a>0$, $b>0$ and $\phi>0$ and is denoted by $\mathcal{TPB}(a,b,\phi)$.
\end{defn}

It can be easily shown by a change of variable $x=1/(y+1)$ that the above density integrates to $1$. The $k$th moment of the TPB distribution is given by
\begin{equation}
\mathbb{E}(X^k)=\frac{\Gamma(a+b)\Gamma(b+k)}{\Gamma(b)\Gamma(a+b+k)} {}_2\mbox{F}_{1}(a+b,b+k;a+b+k;1-\phi)
\end{equation}
where $ {}_2\mbox{F}_{1}$ denotes the hypergeometric function. In fact it can be shown that TPB is a subclass of Gauss hypergeometric (GH) distribution proposed in \cite{armero94} and the compound confluent hypergeometric (CCH) distribution proposed in \cite{gordy98}. 

The density functions of GH and CCH distributions are given by
 \begin{eqnarray}
f_{\small{\mbox{GH}}}(x;a,b,r,\zeta) &=& \frac{x^{b-1}(1-x)^{a-1}(1+\zeta x)^{-r}}{\mbox{B}(b,a){}_2\mbox{F}_{1}(r,b;a+b;-\zeta)}, \label{gh}\\
f_{\small{\mbox{CCH}}}(x;a,b,r,s,\nu,\theta) &=&\frac{\nu^b x^{b-1}(1-x)^{a-1}(\theta+(1-\theta) \nu x)^{-r}}{\mbox{B}(b,a)\exp(-s/\nu)\Phi_1(a,r,a+b,s/\nu,1-\theta)},
\label{cch}
 \end{eqnarray}
for $0<x<1$ and $0<x<1/\nu$, respectively, where $\mbox{B}(b,a)=\Gamma(a)\Gamma(b)/\Gamma(a+b)$ denotes the beta function and $\Phi_1$ is the degenerate hypergeometric function of two variables \cite{gordy98}. Letting $\zeta=\phi-1$, $r=a+b$ and noting that ${}_2\mbox{F}_{1}(a+b,b;a+b;1-\phi)=\phi^{-b}$, (\ref{gh}) becomes a TPB density. Also note that (\ref{cch}) becomes (\ref{gh}) for $s=1$, $\nu=1$ and $\zeta=(1-\theta)/\theta$ \cite{gordy98}. 

\cite{polson2009} considered an alternative special case of the CCH distribution for the shrinkage coefficients, $\rho_j$, by letting $\nu=r=1$ in (\ref{cch}). \cite{polson2009} refer to this special case as the hypergeometric-beta (HB) distribution. TPB and HB generalize the beta distribution in two distinct directions, with one practical advantage of the TPB being that it allows for a straightforward conjugate hierarchy leading to potentially substantial analytical and computational gains.

Now we move onto the hierarchical modeling of a flexible class of shrinkage priors for the estimation of a potentially sparse $p$-vector. Suppose a $p$-dimensional vector $\mathbf{y}|\boldsymbol\theta\sim \mathcal{N}(\boldsymbol\theta,\mathbf{I})$ is observed where $\boldsymbol\theta=(\theta_{1},\ldots,\theta_{p})'$ is of interest. Now we define a shrinkage prior that is obtained by mixing a normal distribution over its scale parameter with the TPB distribution.

\begin{figure}[!t]
\centering \subfigure[]{
\begin{minipage}{.49\linewidth}
 \centering\includegraphics[width=.9\textwidth]{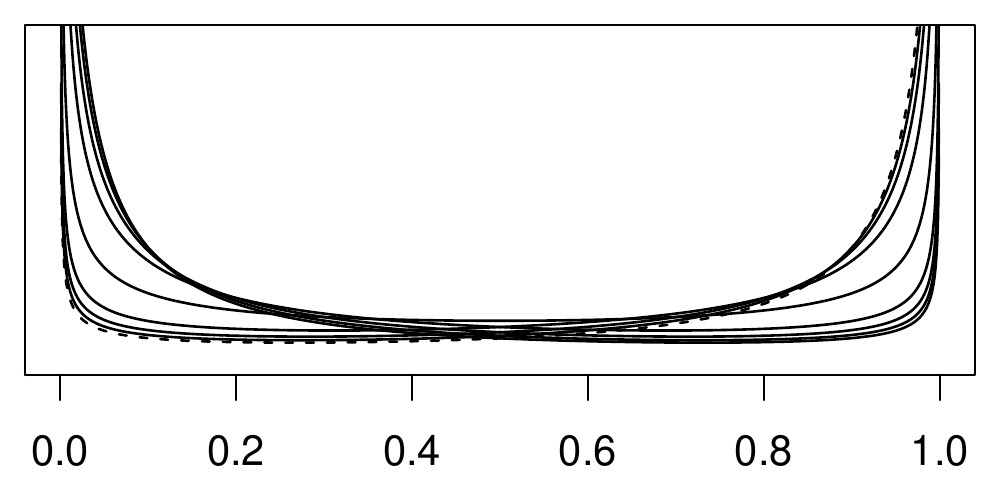}
\end{minipage}}
\centering \subfigure[]{
\begin{minipage}{.49\linewidth}
   \centering\includegraphics[width=.9\textwidth]{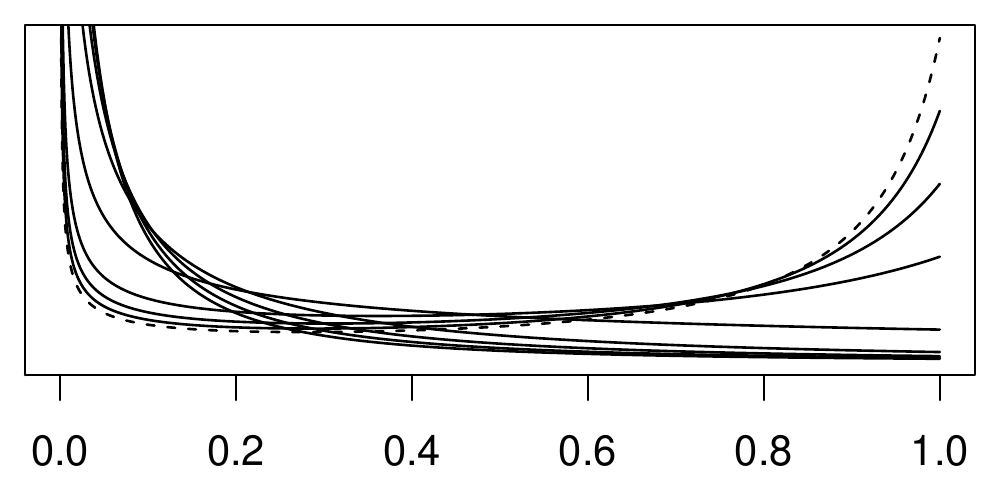}
\end{minipage}}
\centering \subfigure[]{
\begin{minipage}{.49\linewidth}
   \centering\includegraphics[width=.9\textwidth]{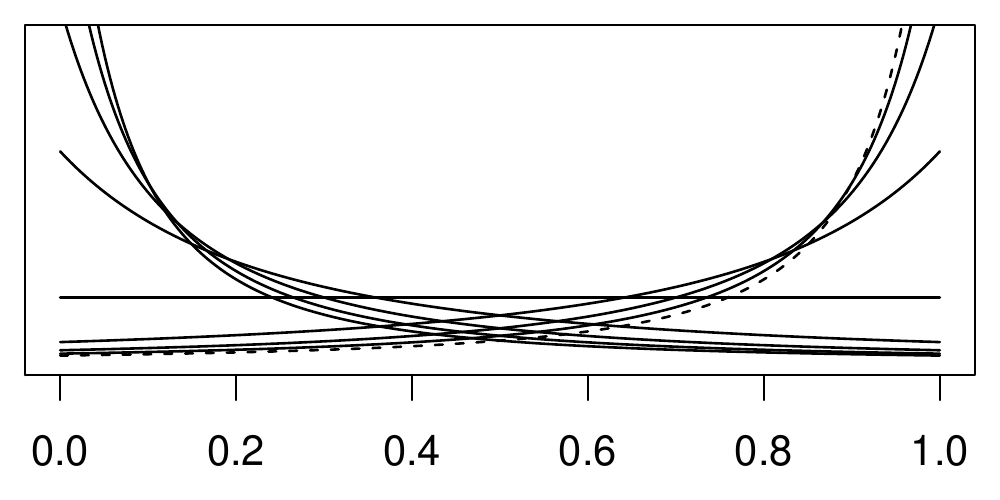}
\end{minipage}}
\centering \subfigure[]{
\begin{minipage}{.49\linewidth}
   \centering\includegraphics[width=.9\textwidth]{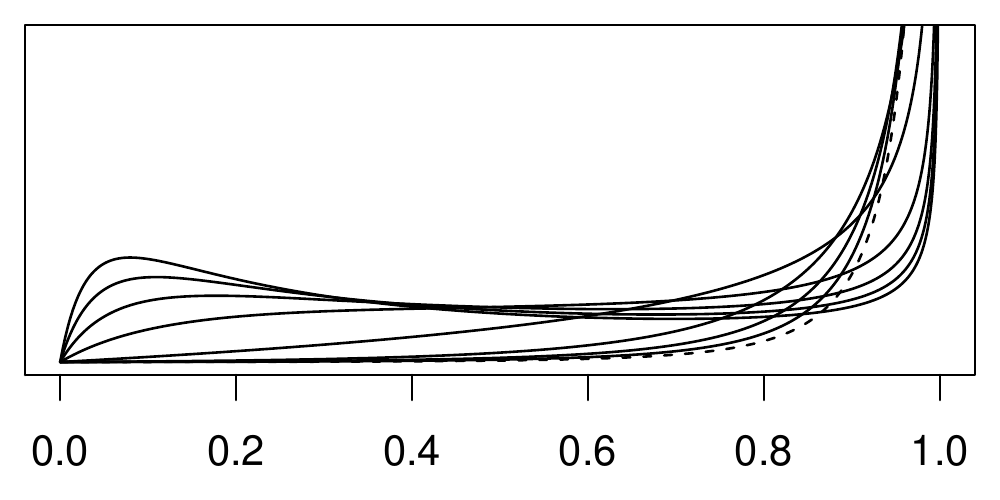}
\end{minipage}}
\centering \subfigure[]{
\begin{minipage}{.49\linewidth}
   \centering\includegraphics[width=.9\textwidth]{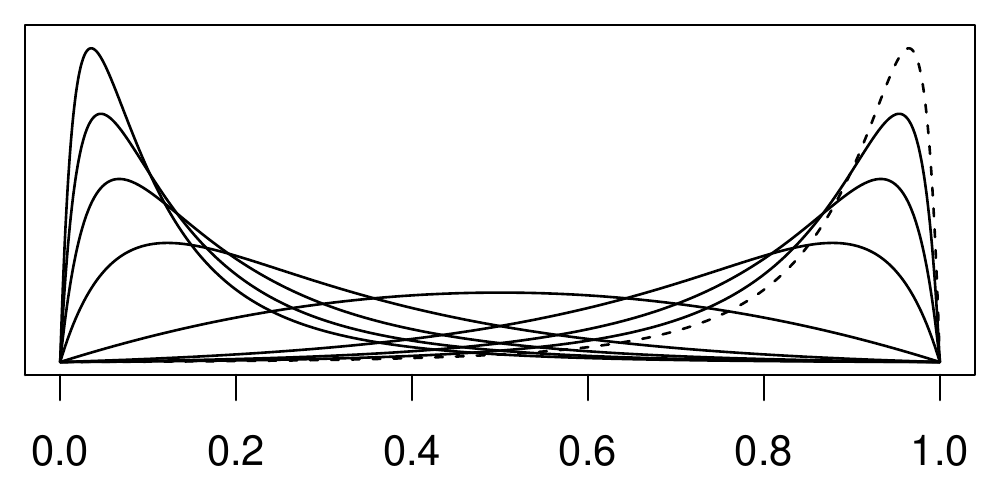}
\end{minipage}}
\centering \subfigure[]{
\begin{minipage}{.49\linewidth}
   \centering\includegraphics[width=.9\textwidth]{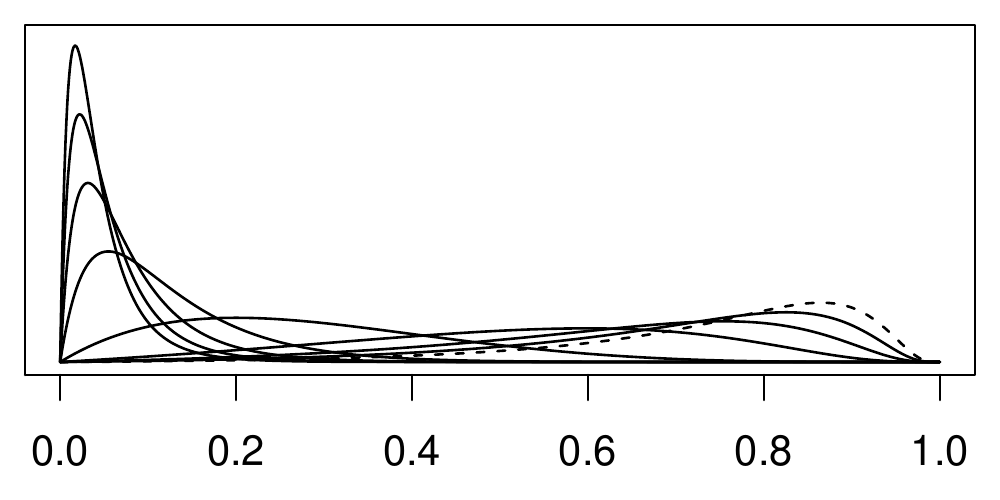}
\end{minipage}}
\caption{$(a,b)=\left\{(1/2,1/2),(1,1/2),(1,1),(1/2,2),(2,2),(5,2)\right\}$ for (a)-(f) respectively. $\phi=\left\{1/10,1/9,1/8,1/7,1/6,1/5,1/4,1/3,1/2,1,2,3,4,5,6,7,8,9,10\right\}$ considered for all pairs of $a$ and $b$. The line corresponding to the lowest value of $\phi$ is drawn with a dashed line.\label{fig1}} 
\end{figure}

\begin{defn}
The TPB normal scale mixture representation for the distribution of random variable $\theta_j$ is given by
\begin{equation}
\theta_j|\rho_j\sim \mathcal{N}(0,1/\rho_j-1), \ \ \ \rho_j\sim \mathcal{TPB}(a,b,\phi),
\label{betalike}
\end{equation}
where $a>0$, $b>0$ and $\phi>0$. The resulting marginal distribution on $\theta_j$ is denoted by $\mathcal{TPBN}(a,b,\phi)$.
\end{defn}
Figure \ref{fig1} illustrates the density on $\rho_j$ for varying values of $a$, $b$ and $\phi$. Note that the special case for $a=b=1/2$ in Figure \ref{fig1}(a) gives the horseshoe prior. Also when $a=\phi=1$ and $b=1/2$, this representation yields the Strawderman-Berger prior. For a fixed value of $\phi$, smaller $a$ values yield a density on $\theta_{j}$ that is more peaked at zero, while smaller values of $b$ yield a density on $\theta_{j}$ that is heavier tailed. For fixed values of $a$ and $b$, decreasing $\phi$ shifts the mass of the density on $\rho_{j}$ from left to right, suggesting more support for stronger shrinkage. That said, the density assigned in the neighborhood of $\theta_{j}=0$ increases while making the overall density lighter-tailed. We next propose the equivalence of three hierarchical representations revealing a wide class of priors encompassing many of those mentioned earlier.

\begin{prop}
If $\theta_j\sim \mathcal{TPBN}(a,b,\phi)$, then\\
1) $\theta_{j}\sim \mathcal{N}(0,\tau_{j})$, $\tau_{j} \sim \mathcal{G}(a,\lambda_{j})$ and $\lambda_{j} \sim \mathcal{G}(b,\phi)$. \\
2) $\theta_{j}\sim \mathcal{N}(0,\tau_{j})$, $\pi(\tau_{j})=\frac{\Gamma(a+b)}{\Gamma(a)\Gamma(b)}\phi^{-a}\tau^{a-1}(1+\tau_{j}/\phi)^{-(a+b)}$ which implies that $\tau_{j}/\phi\sim\beta'(a,b)$, the inverted beta distribution with parameters $a$ and $b$.
\end{prop}

The equivalence given in Proposition 1 is significant as it makes the work in Section 4 possible under the TPB normal scale mixtures as well as further revealing connections among previously proposed shrinkage priors. It provides a rich class of priors leading to great flexibility in terms of the induced shrinkage and makes it clear that this new class of priors can be considered simultaneous extensions to the work by \cite{griffin2007,griffin2010} and \cite{Carvalho2009,carvalho2010}. It is worth mentioning that the hierarchical prior(s) given in Proposition 1 are different than the approach taken by \cite{griffin2010} in how we handle the mixing. In particular, the first hierarchy presented in Proposition 1 is identical to the NG prior up to the first stage mixing. While fixing the values of $a$ and $b$, we further mix over $\lambda_{j}$ (rather than a global $\lambda$) and further over $\phi$ if desired as will be discussed later. $\phi$ acts as a global shrinkage parameter in the hierarchy. On the other hand, \cite{griffin2010} choose to further mix over $a$ and a global $\lambda$ while fixing the values of $b$ and $\phi$. By doing so, they forfeit a complete conjugate structure and an explicit control over the tail behavior of $\pi(\theta_j)$.

As a direct corollary to Proposition 1, we observe a possible equivalence between the SB and the NEG priors.
\begin{cor} If $a=1$ in Proposition 1, then $\mbox{TPBN}\equiv \mbox{NEG}$. If $(a,b,\phi)=(1,1/2,1)$ in Proposition 1, then $\mbox{TPBN}\equiv \mbox{SB}\equiv \mbox{NEG}$.
\end{cor}

An interesting, yet expected, observation on Proposition 1 is that a half-Cauchy prior can be represented as a scale mixture of gamma distributions, i.e. if $\tau_{j} \sim \mathcal{G}(1/2,\lambda_{j})$ and $\lambda_{j} \sim \mathcal{G}(1/2,\phi)$, then $\tau_{j}^{1/2}\sim \mathcal{C}^{+}(0,\phi^{1/2})$. This makes sense as $\tau^{1/2}|\lambda_j$ has a half-Normal distribution and the mixing distribution on the precision parameter is gamma with shape parameter $1/2$.

\cite{carvalho2010} further place a half-Cauchy prior on $\phi^{1/2}$ to complete the hierarchy. The aforementioned observation helps us formulate the complete hierarchy proposed in \cite{carvalho2010} in a conjugate manner. This should bring analytical and computational advantages as well as making the application of the procedure much easier for the average user without the need for a relatively more complex sampling scheme.

\begin{cor}
If $\theta_{j}\sim \mathcal{N}(0,\tau_{j})$, $\tau_{j}^{1/2}\sim \mathcal{C}^{+}(0,\phi^{1/2})$ and $\phi^{1/2}\sim\mathcal{C}^{+}(0,1)$, then $\theta_j\sim\mathcal{TPBN}(1/2,1/2,\phi)$, $\phi\sim\mathcal{G}(1/2,\omega)$ and $\omega\sim\mathcal{G}(1/2,1)$.
\end{cor}

Hence disregarding the different treatments of the higher-level hyper-parameters, we have shown that the class of priors given in Definition 2 unites the priors in \cite{johnstone2004,griffin2007,Carvalho2009,carvalho2010} under one family and reveals their close connections through the equivalence of hierarchies given in Proposition 1. The first hierarchy in Proposition 1 makes much of the work possible in the following sections. 

\section{Estimation and Posterior Inference in Regression Models}
\subsection{Fully Bayes and Approximate Inference}
Consider the linear regression model,
$\mathbf{y}=\mathbf{X}{\boldsymbol\beta}+{\boldsymbol\epsilon}$,
where $\mathbf{y}$ is an $n$-dimensional vector of responses,
$\mathbf{X}$ is the $n\times p$ design matrix and
$\boldsymbol\epsilon$ is an $n$-dimensional vector of independent residuals which are normally distributed,
$\mathcal{N}(\mathbf{0},\sigma^{2}\mathbf{I}_{n})$
with variance $\sigma^{2}$.

We place the hierarchical prior given in Proposition 1 on each $\beta_{j}$, i.e. $\beta_{j} \sim \mathcal{N}(0,\sigma^{2}\tau_{j})$, $\tau_{j} \sim \mathcal{G}(a,\lambda_{j})$, $\lambda_{j} \sim \mathcal{G}(b,\phi)$. $\phi$ is used as a global shrinkage parameter common to all $\beta_{j}$, and may be inferred using the data. Thus we follow the hierarchy by letting $\phi \sim \mathcal{G}(1/2,\omega)$, $\omega \sim \mathcal{G}(1/2,1)$ which implies $\phi^{1/2}\sim \mathcal{C}^{+}(0,1)$ that is identical to what was used in \cite{carvalho2010} at this level of the hierarchy. However, we do not believe at this level in the hierarchy the choice of the prior will have a huge impact on the results. Although treating $\phi$ as unknown may be reasonable, when there exists some prior knowledge, it is appropriate to fix a $\phi$ value to reflect our prior belief in terms of underlying sparsity of the coefficient vector. This sounds rather natural as soon as one starts seeing $\phi$ as a parameter that governs the multiplicity adjustment as discussed in \cite{carvalho2010}. Note also that here we form the dependence on the error variance at a lower level of hierarchy rather than forming it in the prior of $\phi$ as done in \cite{carvalho2010}. If we let $a=b=1/2$, we will have formulated the hierarchical prior given in \cite{carvalho2010} in a completely conjugate manner. We also let $\sigma^{-2}\sim \mathcal{G}(c_{0}/2,d_{0}/2)$. Under a normal likelihood, an efficient Gibbs sampler may be obtained as the fully conditional posteriors can be extracted: $\boldsymbol\beta|\mathbf{y},\mathbf{X},\sigma^{2},\tau_{1},\ldots,\tau_{p}\sim \mathcal{N}(\boldsymbol\mu_{\beta},\mathbf{V}_{\beta})$, $\sigma^{-2}|\mathbf{y},\mathbf{X},\boldsymbol\beta,\tau_{1},\ldots,\tau_{p}\sim \mathcal{G}(c^{*},d^{*})$, $\tau_{j}|\beta_{j},\sigma^{2},\lambda_{j}\sim \mathcal{GIG}(a-1/2,2\lambda_{j},\beta_{j}^{2}/\sigma^{2})$, $\lambda_{j}|\tau_{j},\phi\sim \mathcal{G}(a+b,\tau_{j}+\phi)$, $\phi|\lambda_{j},\omega\sim \mathcal{G}(pb+1/2,\sum_{j=1}^{p}\lambda_{j}+\omega)$, $\omega|\phi\sim \mathcal{G}(1,\phi+1)$, where $\boldsymbol\mu_{\beta}=(\mathbf{X}'\mathbf{X}+\mathbf{T}^{-1})^{-1}\mathbf{X}'\mathbf{y}$, $\mathbf{V}_{\beta}=\sigma^{2}(\mathbf{X}'\mathbf{X}+\mathbf{T}^{-1})^{-1}$, $c^{*}=(n+p+c_{0})/2$, $d^{*}=\{(\mathbf{y}-\mathbf{X}\boldsymbol\beta)'(\mathbf{y}-\mathbf{X}\boldsymbol\beta)+\boldsymbol\beta'\mathbf{T}^{-1}\boldsymbol\beta+d_{0}\}/2$, $\mathbf{T}=\mbox{diag}(\tau_{1},\ldots,\tau_{p})$.

As an alternative to MCMC and Laplace approximations \cite{tierney1986}, a lower-bound on marginal likelihoods may be obtained via variational methods \cite{jordan99} yielding approximate posterior distributions on the model parameters. Using a similar approach to  \cite{bishop01,armagan09a}, the approximate marginal posterior distributions of the parameters are given by 
$\boldsymbol\beta\sim\mathcal{N}(\boldsymbol\mu_{\beta},\mathbf{V}_{\beta})$, $\sigma^{-2}\sim\mathcal{G}\left(c^{*},d^{*}\right)$, $\tau_{j}\sim\mathcal{GIG}(a-1/2,2\langle\lambda_{j}\rangle,\langle\sigma^{-2}\rangle\langle\beta_{j}^{2}\rangle)$, $\lambda_{j}\sim\mathcal{G}(a+b,\langle\tau_{j}\rangle+\langle\phi\rangle)$, $\phi\sim\mathcal{G}(pb+1/2,\langle\omega\rangle+\sum_{j=1}^{p}\langle\lambda_{j}\rangle)$, $\omega\sim\mbox{G}(1,\langle\phi\rangle+1)$, where $\boldsymbol\mu_{\beta} = \langle\boldsymbol\beta\rangle = (\mathbf{X}'\mathbf{X}+\mathbf{T}^{-1})^{-1}\mathbf{X}'\mathbf{y}$, $\mathbf{V}_{\beta} = \langle\sigma^{-2}\rangle^{-1}(\mathbf{X}'\mathbf{X}+\mathbf{T}^{-1})^{-1}$, $\mathbf{T}^{-1}=\mbox{diag}(\langle\tau_{1}^{-1}\rangle,\ldots,\langle\tau_{p}^{-1}\rangle)$, $c^{*} = (n+p+c_{0})/2$, $d^{*} = (\mathbf{y}'\mathbf{y}-2\mathbf{y}'\mathbf{X}\langle\boldsymbol\beta\rangle+\sum_{i=1}^{n}\mathbf{x}_{i}\langle\boldsymbol\beta\boldsymbol\beta'\rangle\mathbf{x}_{i}+\sum_{j=1}^{p}\langle\beta_{j}^{2}\rangle\langle\tau_{j}^{-1}\rangle+d_{0})/2$, $\langle\boldsymbol\beta\boldsymbol\beta'\rangle=\mathbf{V}_{\beta}+\langle\boldsymbol\beta\rangle\langle\boldsymbol\beta'\rangle$, $\langle\sigma^{-2}\rangle = c^{*}/d^{*}$, $\langle\lambda_{j}\rangle=(a+b)/(\langle\tau_{j}\rangle+\langle\phi\rangle)$, $\langle\phi\rangle=(pb+1/2)/(\langle\omega\rangle+\sum_{j=1}^{p}\langle\lambda_{j}\rangle)$, $\langle\omega\rangle=1/(\langle\phi\rangle+1)$ and
\begin{eqnarray}
\langle\tau\rangle&=&\frac{(\langle\sigma^{-2}\rangle\langle\beta_{j}^{2}\rangle)^{1/2} \mbox{K}_{a+1/2}\left\{(2\langle\lambda_{j}\rangle\langle\sigma^{-2}\rangle\langle\beta_{j}^{2}\rangle)^{1/2}\right\}}{(2\langle\lambda_{j}\rangle)^{1/2}\mbox{K}_{a-1/2}\left\{(2\langle\lambda_{j}\rangle\langle\sigma^{-2}\rangle\langle\beta_{j}^{2}\rangle)^{1/2}\right\}},\nonumber\\
\langle\tau^{-1}\rangle&=&\frac{(2\langle\lambda_{j}\rangle)^{1/2}\mbox{K}_{3/2-a}\left\{(2\langle\lambda_{j}\rangle\langle\sigma^{-2}\rangle\langle\beta_{j}^{2}\rangle)^{1/2}\right\}}{(\langle\sigma^{-2}\rangle\langle\beta_{j}^{2}\rangle)^{1/2}\mbox{K}_{1/2-a}\left\{(2\langle\lambda_{j}\rangle\langle\sigma^{-2}\rangle\langle\beta_{j}^{2}\rangle)^{1/2}\right\}}.\nonumber
\end{eqnarray}
This procedure consists of initializing the moments and iterating through them until some convergence criterion is reached. The deterministic nature of these approximations make them attractive as a quick alternative to MCMC. 

This conjugate modeling approach we have taken allows for a very straightforward implementation of Strawderman-Berger and horseshoe priors or, more generally, TPB normal scale mixture priors in regression models without the need for a more sophisticated sampling scheme which may ultimately attract more audiences towards the use of these more flexible and carefully defined normal scale mixture priors.

\subsection{Sparse Maximum a Posteriori Estimation}
\begin{figure}[!t]
\centering \subfigure[]{
\begin{minipage}{.49\linewidth}
 \centering\includegraphics[width=1\textwidth]{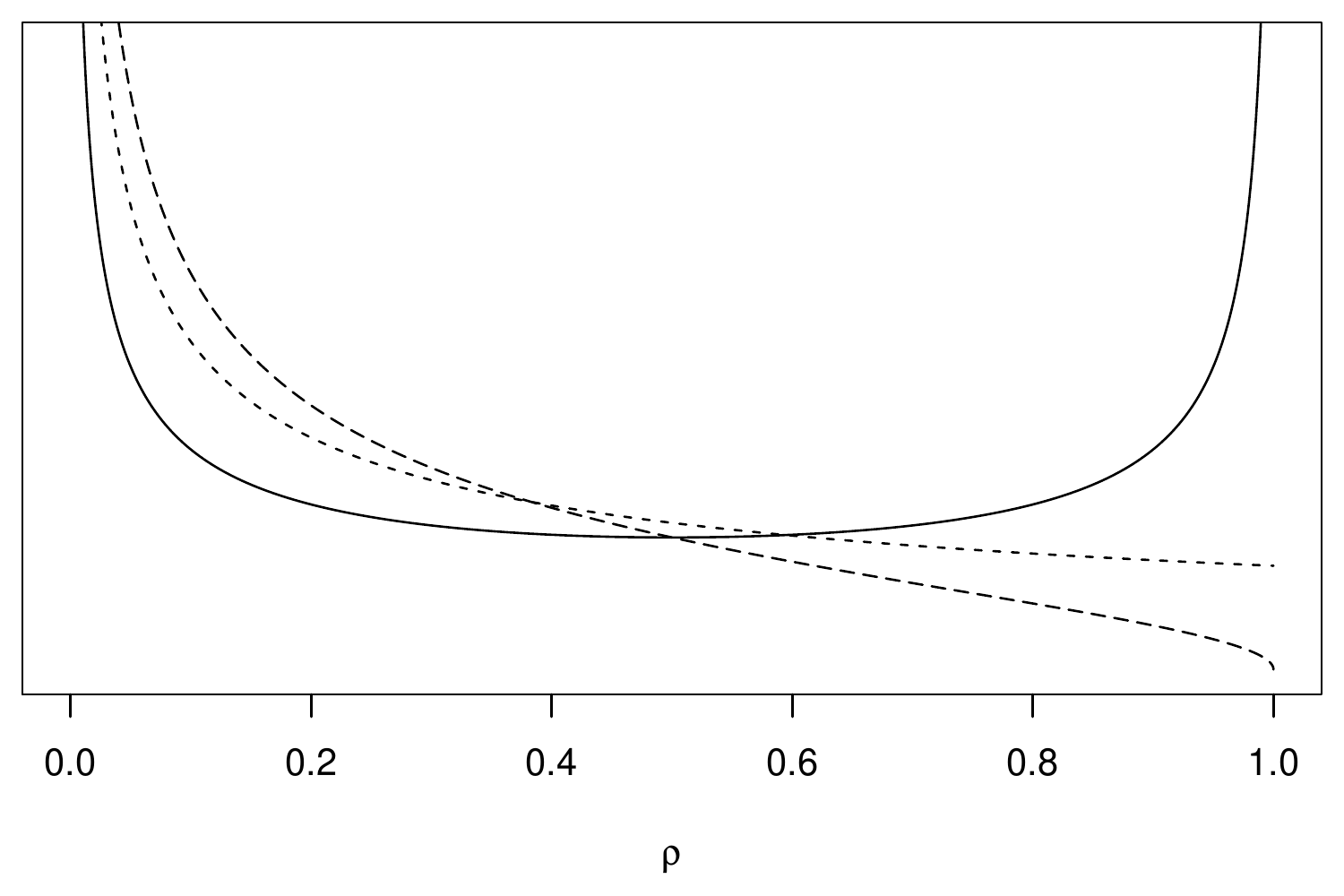}
\end{minipage}}
\centering \subfigure[]{
\begin{minipage}{.49\linewidth}
   \centering\includegraphics[width=1\textwidth]{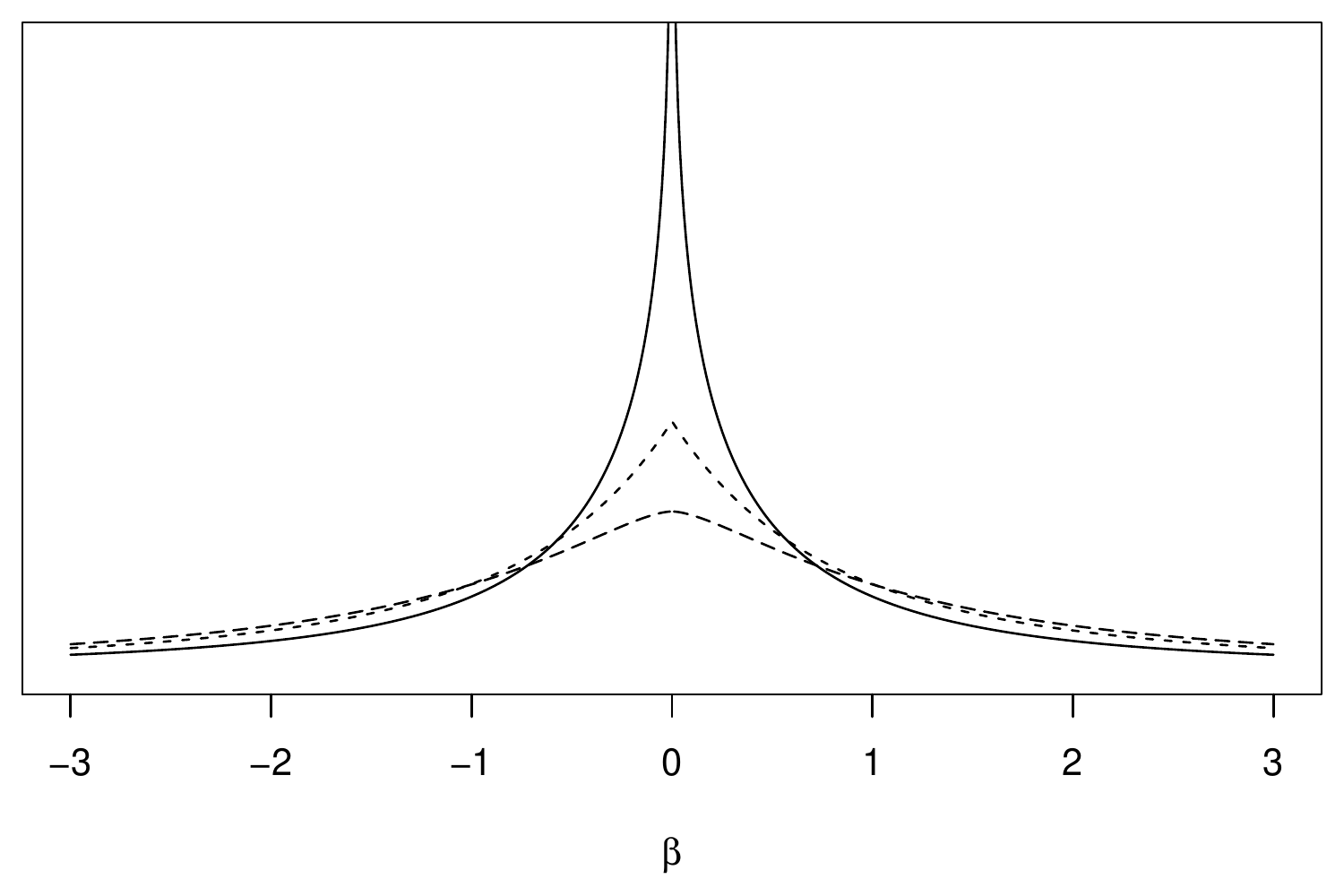}
\end{minipage}}
\caption{Prior densities of (a) $\rho_{j}$ and (b) $\beta_{j}$ for $a=1/2$ (solid), $a=1$ (dashed) and $a=3/2$ (long dash).  \label{fig2}} 
\end{figure}

Although not our main focus, many readers are interested in sparse solutions, hence we give the following brief discussion. Given $a$, $b$ and $\phi$, maximum a posteriori (MAP) estimation is rather straightforward via a simple expectation-maximization (EM) procedure. This is accomplished in a similar manner to \cite{figo03} by obtaining the joint MAP estimates of the error variance and the regression coefficients having taken the expectation with respect to the conditional posterior distribution of $\tau_{j}^{-1}$ using the second hierarchy given in Proposition 1. The $k$th expectation step then would consist of calculating
\begin{equation}
\langle\tau_{j}^{-1}\rangle^{(k)}=\frac{\int_{0}^{\infty}\tau_j^{a-1/2}(1+\tau_j/\phi)^{-(a+b)}\exp\{-\beta_{j}^{2(k-1)}/(2\sigma^{2}_{(k-1)}\tau_j)\}d\tau_{j}^{-1}}{\int_{0}^{\infty}\tau_j^{1/2+a}(1+\tau_j/\phi)^{-(a+b)}\exp\{-\beta_{j}^{2(k-1)}/(2\sigma^{2}_{(k-1)}\tau_j)\}d\tau_{j}^{-1}}
\label{expectation}
\end{equation}
where $\beta_{j}^{2(k-1)}$ and $\sigma^{2}_{(k-1)}$ denote the modal estimates of the $j$th component of $\boldsymbol\beta$ and the error variance $\sigma^2$ at iteration $(k-1)$.
The solution to (\ref{expectation}) may be expressed in terms of some special function(s) for changing values of $a$, $b$ and $\phi$. $b<1$ is a good choice as it will keep the tails of the marginal density on $\beta_{j}$ heavy. A careful choice of $a$, on the other hand, is essential to sparse estimation. Admissible values of $a$ for sparse estimation is apparent by the representation in Definition 2, noting that for any $a>1$, $\pi(\rho_{j}=1)=0$, i.e. $\beta_{j}$ may never be shrunk exactly to zero. Hence for sparse estimation, it is essential that $0<a\leq1$. Figure \ref{fig2} (a) and (b) give the prior densities on $\rho_{j}$ for $b=1/2$, $\phi=1$ and $a=\{1/2,1,3/2\}$ and the resulting marginal prior densities on $\beta_{j}$. These marginal densities are given by
\begin{equation} 
\pi(\beta_{j})=\left\{ \begin{array}{ll} 
\frac{1}{\sqrt{2}\pi^{3/2}}e^{\beta_j^2/2}\Gamma(0,\beta_j^2/2) & a=1/2\\
\frac{1}{\sqrt{2\pi}}-\frac{|\beta_j|}{2}e^{\beta_j^2/2}+\frac{\beta_{j}}{2}e^{\beta_j^2/2}\mbox{Erf}(\beta_{j}/\sqrt{2}) & a=1\\
\frac{\sqrt{2}}{\pi^{3/2}}\left\{1-\frac{1}{2}e^{\beta_{j}^2/2}\beta_{j}^2\Gamma(0,\beta_{j}^2/2)\right\} & a=3/2
				\end{array} \right.\nonumber
\end{equation}
where $\mbox{Erf}(.)$ denotes the error function and $\Gamma(s,z)=\int_{z}^{\infty}t^{s-1}e^{-t}dt$ is the incomplete gamma function. Figure \ref{fig2} clearly illustrates that while all three cases have very similar tail behavior, their behavior around the origin differ drastically.

\section{Experiments}
\begin{figure}[!t]
\centering \subfigure[]{
\begin{minipage}{.48\linewidth}
 \centering\includegraphics[width=1\textwidth]{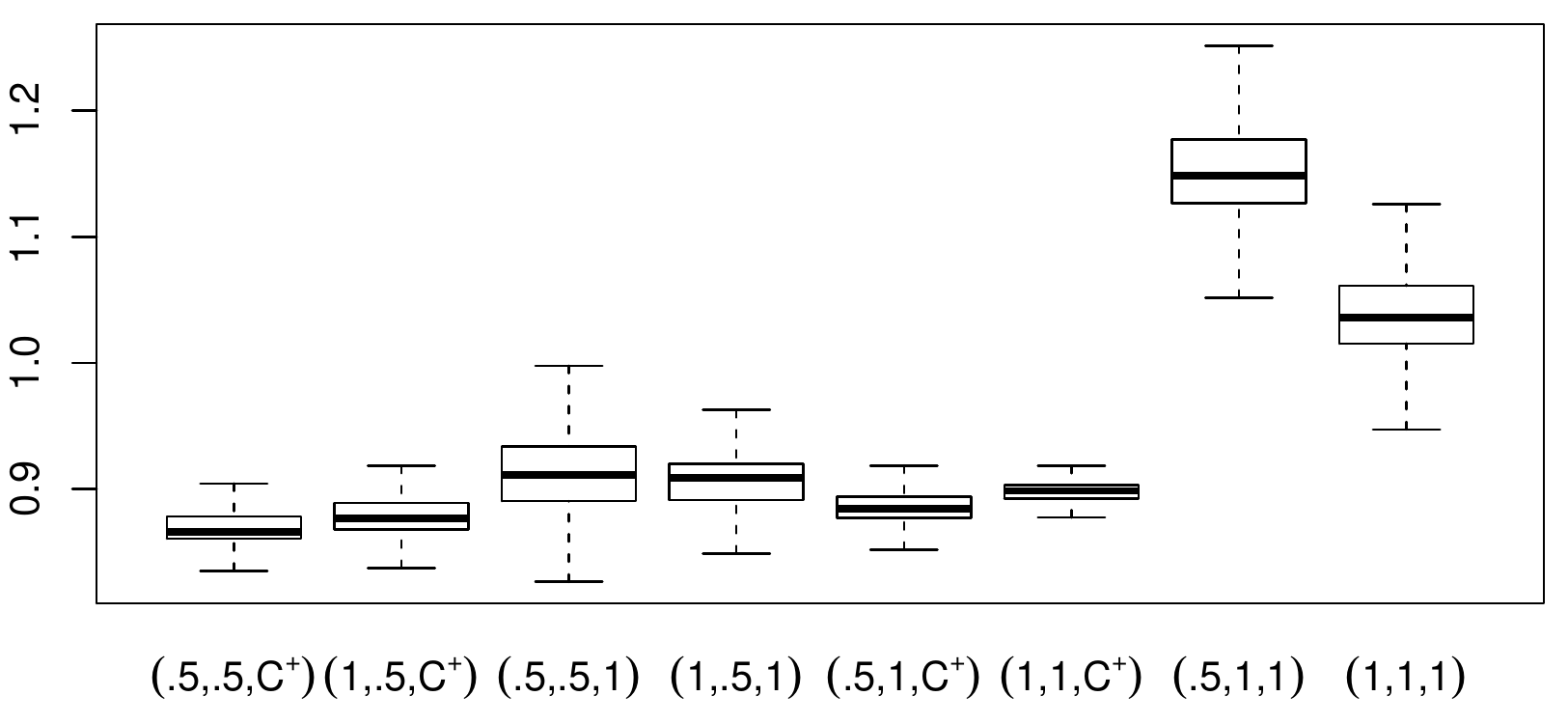} \\ 
\end{minipage}}
\centering \subfigure[]{
\begin{minipage}{.48\linewidth}
 \centering\includegraphics[width=1\textwidth]{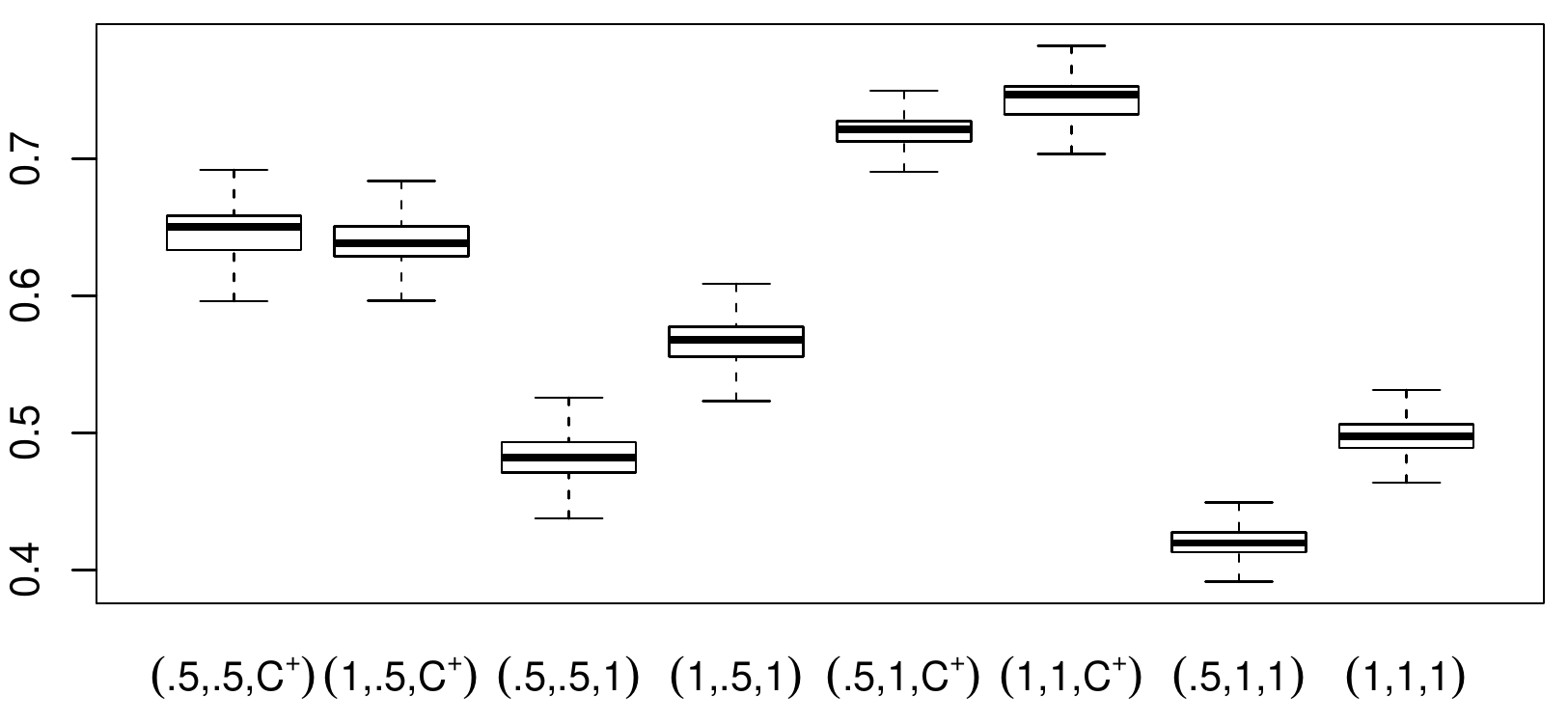} \\ 
\end{minipage}}
\caption{Relative ME at different $(a,b,\phi)$ values for (a) Case 1 and (b) Case 2.\label{fig3}}
\end{figure}

Throughout this section we use the Jeffreys' prior on the error precision by setting $c_0=d_0=0$. We generate data for two cases, $(n,p)=\{(50,20),(250,100)\}$, from $y_{i}=\mathbf{x}'_{i}\boldsymbol\beta^{*}+\epsilon_{i}$, for $i=1,\ldots,n$ where $\boldsymbol\beta^{*}$ is a $p$-vector that on average contains $20q$ non-zero elements which are indexed by the set $\mathcal{A}=\{j: \beta_{j}^{*}\neq 0\}$ for some random $q\in(0,1)$. We randomize the procedure in the following manner: (i) $\mathbf{C} \sim \mathcal{W}(p,\mathbf{I}_{p\times p})$, (ii) $\mathbf{x}_{i} \sim \mathcal{N}(\mathbf{0},\mathbf{C})$, (iii) $q \sim \mathcal{B}(1,1)$ for the first and $q \sim \mathcal{B}(1,4)$ for the second cases, (iv) $I(j\in \mathcal{A}) \sim \mbox{Bernoulli}(q)$ for $j=1,\ldots,p$ where $I(.)$ denotes the indicator function, (v) for $j\in\mathcal{A}$, $\beta_{j}\sim \mathcal{U}(0,6)$ and for $j\notin\mathcal{A}$, $\beta_{j}=0$ and finally (vi) $\epsilon_{i} \sim \mathcal{N}(0,\sigma^{2})$ where $\sigma \sim \mathcal{U}(0,6)$. We generated $1000$ data sets for each case resulting in a median signal-to-noise ratio of approximately $3.3$ and $4.5$. We obtain the estimate of the regression coefficients, $\hat{\boldsymbol\beta}$, using the variational Bayes procedure and measure the performance by model error which is calculated as $(\boldsymbol\beta^*-\hat{\boldsymbol\beta})'\mathbf{C}(\boldsymbol\beta^*-\hat{\boldsymbol\beta})$. Figure \ref{fig3}(a) and (b) display the median \emph{relative} model error (RME) values (with their distributions obtained via bootstrapping) which is obtained by dividing the model error observed from our procedures by that of $\ell_1$ regularization (lasso) tuned by $10$-fold cross-validation. The boxplots in Figure \ref{fig3}(a) and (b) correspond to different $(a,b,\phi)$ values where $\mbox{C}^+$ signifies that $\phi$ is treated as unknown with a half-Cauchy prior as given earlier in Section 4.1. It is worth mentioning that we attain a clearly superior performance compared to the lasso, particularly in the second case, despite the fact that the estimator resulting from the variational Bayes procedure is not a thresholding rule. Note that $b=1$ choice leads to much better performance under Case 2 than Case 1. This is due to the fact that Case 2 involves a much sparser underlying setup on average than Case 1 and that the lighter tails attained by setting $b=1$ leads to stronger shrinkage. 

\begin{figure}[!t]
\centering\includegraphics[width=1\textwidth]{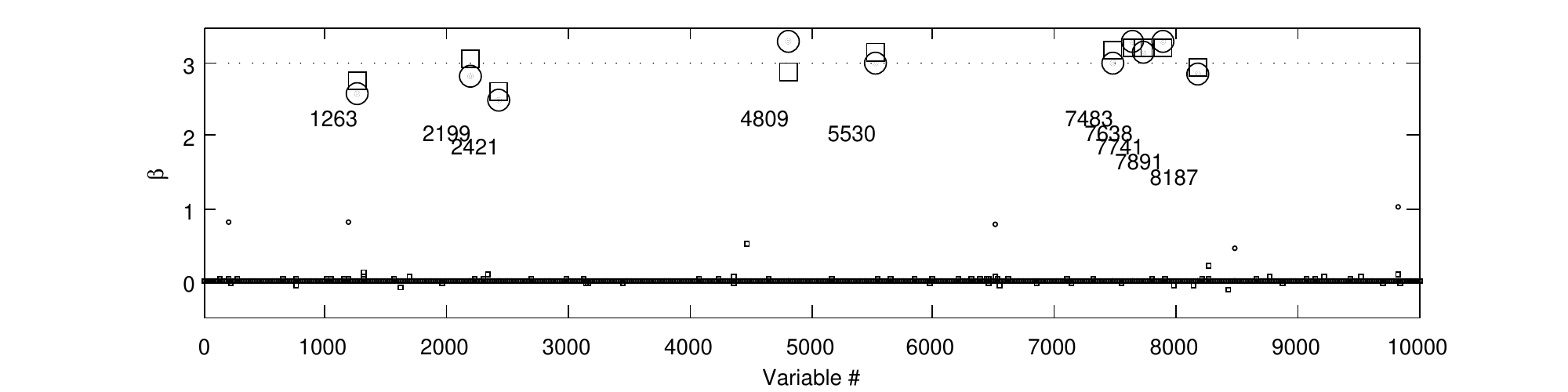} \\ 
\caption{Posterior mean of $\boldsymbol\beta$ by sampling (square) and by approximate inference (circle).\label{fig4}}
\end{figure}

To give a high dimensional example, we also generate a data set from the model $y_{i}=\mathbf{x}'_{i}\boldsymbol\beta^{*}+\epsilon_{i}$, for $i=1,\ldots,100$, where $\boldsymbol\beta^{*}$ is a $10000$-dimensional very sparse vector with $10$ randomly chosen components set to be $3$, $\epsilon_{i}\sim \mathcal{N}(0,3^2)$ and $x_{ij}\sim\mathcal{N}(0,1)$ for $j=1,\dots,p$. This $\boldsymbol\beta^{*}$ choice leads to a signal-to-noise ratios of $3.16$. For the particular data set we generated, the randomly chosen components of $\boldsymbol\beta^{*}$ to be non-zero were indexed by $1263$, $2199$, $2421$, $4809$, $5530$, $7483$, $7638$, $7741$, $7891$ and $8187$. We set $(a,b,\phi)=(1,1/2,10^{-4})$ which implies that a priori $\mathbb{P}(\rho_{j}>0.5)=0.99$ placing much more density in the neighborhood of $\rho_j=1$ (total shrinkage). This choice is due to the fact that $n/p=0.01$ and to roughly reflect that we do not want any more than $100$ predictors in the resulting model. Hence $\phi$ is used, a priori, to limit the number of predictors in the model in relation to the sample size. Also note that with $a=1$, the conditional posterior distribution of $\tau_{j}^{-1}$ is reduced to an inverse Gaussian. Since we are adjusting the global shrinkage parameter, $\phi$, a priori, and it is chosen such that $\mathbb{P}(\rho_{j}>0.5)=0.99$, whether $a=1/2$ or $a=1$ should not matter. We first run the Gibbs sampler for $100000$ iterations ($2.4$ hours on a computer with a $2.8$ GHz CPU and $12$ Gb of RAM using \texttt{Matlab}), discard the first $20000$, thin the rest by picking every $5$th sample to obtain the posteriors of the parameters. We observed that the chain converged by the $10000$th iteration. For comparison purposes, we also ran the variational Bayes procedure using the values from the converged chain as the initial points ($80$ seconds). Figure \ref{fig4} gives the posterior means attained by sampling and the variational approximation. The estimates corresponding to the zero elements of $\boldsymbol\beta^{*}$ are plotted with smaller shapes to prevent clutter. We see that in both cases the procedure is able to pick up the larger signals and shrink a significantly large portion of the rest towards zero. The approximate inference results are in accordance with the results from the Gibbs sampler. It should be noted that using a good informed guess on $\phi$, rather than treating it as an unknown in this high dimensional setting, improves the performance drastically.

\section{Discussion}
We conclude that the proposed hierarchical prior formulation constitutes a useful encompassing framework in understanding the behavior of different scale mixtures of normals and connecting them under a broader family of hierarchical priors. While $\ell_1$ regularization, or namely lasso, arising from a double exponential prior in the Bayesian framework yields certain computational advantages, it demonstrates much inferior estimation performance relative to the more carefully formulated scale mixtures of normals. The proposed equivalence of the hierarchies in Proposition 1 makes computation much easier for the $\mbox{TPB}$ scale mixtures of normals. As per different choices of hyper-parameters, we recommend that $a \in (0,1]$ and $b \in (0,1)$; in particular $(a,b)=\{(1/2,1/2),(1,1/2)\}$. These choices guarantee that the resulting prior has a kink at zero, which is essential for sparse estimation, and leads to heavy tails to avoid unnecessary bias in large signals (recall that a choice of $b=1/2$ will yield Cauchy-like tails). In problems where oracle knowledge on sparsity exists or when $p>>n$, we recommend that $\phi$ is fixed at a reasonable quantity to reflect an appropriate sparsity constraint as mentioned in Section 5. 

\subsubsection*{Acknowledgments}
This work was supported by Award Number R01ES017436 from the National Institute of Environmental Health Sciences.  The content is solely the responsibility of the authors and does not necessarily represent the official views of the National Institute of Environmental Health Sciences or the National Institutes of Health.

\subsubsection*{References}
\vspace{-20pt}
\bibliography{paper_v15}
\bibliographystyle{plain}

\end{document}